\documentstyle[12pt,aasms4,psfig]{article}
\lefthead{Ota et al.}
\righthead{Galaxy Cluster at $z=0.62$}

\begin{document}
\title{
 Detection of Iron Emission Line from the Galaxy Cluster 
Including the Radio Galaxy 3C220.1 at $z=0.62$
}
\author{
Naomi Ota (ota@astro.isas.ac.jp)\altaffilmark{1},\\
Kazuhisa Mitsuda (mitsuda@astro.isas.ac.jp)\altaffilmark{1},\\
Makoto Hattori (hattori@astr.tohoku.ac.jp)\altaffilmark{2},\\
and\\ 
Tatehiro Mihara (mihara@crcosmo.riken.go.jp)\altaffilmark{3}\\
}
\altaffiltext{1}
{Institute of Space and Astronautical Science, 
3-1-1 Yoshinodai, Sagamihara, Kanagawa, 229-8510, Japan}
\altaffiltext{2}
{Astronomical Institute, T\^{o}hoku University, Aoba, 
Sendai, 980-8578, Japan}
\altaffiltext{3}
{Institute of Physical and Chemical Research, 2-1 Hirosawa, Wako, 
Saitama, 351-0198, Japan}

\begin{abstract} 
We have detected an emission line feature at 4 keV in the X-ray
emission from a sky region including the distant radio galaxy
3C220.1($z=0.62$) obtained with ASCA.  The line energy is 6.1 -- 7.0
keV (90\% confidence) in the rest frame of 3C220.1. Within the present
statistics, the observed spectra are consistent with two different
models: a non-thermal model consisting of a power-law continuum plus a
6.4 keV iron emission line, and a Raymond-Smith thin-thermal emission
model of k$T \sim 6$ keV with a metal abundance of $\sim0.5$
solar. However, because of the large ($\sim 500$ eV) equivalent width
of the line, a significant fraction of the X-ray emission is likely to
arise from the hot intracluster gas associated with the galaxy
cluster that includes 3C220.1. The spectral parameters of the thermal
emission are consistent with the luminosity-temperature relation of
nearby clusters.
\end{abstract}

\keywords{Galaxies:Clusters:Individual (3C220.1) --
          X-Rays:Galaxies --  Cosmology:Gravitational Lensing}

\section{Introduction} 
Among a variety of attempts searching for clusters at high redshifts,
X-ray observation directly illuminates the gravitational potential
well because the X-ray-emitting hot plasma is considered to trace
it. Moreover, X-ray emission lines from highly ionized ions, iron in
particular, can be used to determine the redshift of the hot gas, and
thereby its association with other objects can be investigated. Based
on these ideas we have conducted an observation of the radio galaxy
3C220.1 with ASCA.

The existence of a galaxy cluster which surrounds the radio galaxy
3C220.1 ($z=0.62$) was first suggested from the observations at Lick
Observatory (\cite{DICKINSON.1984}). Around the radio galaxy in the
Lick image, several quite red galaxies were found. Further
observations were performed at Kitt Peak National Observatory, and the
blue band image revealed the presence of a giant luminous arc
(\cite{DICKINSON.1984}).  In 1995, much higher quality images were
obtained with the Hubble Space Telescope (\cite{DICKINSON.1998}). A large
arc (9 arc seconds in radius and subtending $\sim$ 70 degrees around
the radio galaxy) was clearly resolved.

The arc image of 3C220.1 is regarded as a section
of an Einstein ring caused by gravitational lensing. This offers a
remarkable tool to constrain the cluster mass enclosed within it. The
redshift of the arc was successfully determined using the Keck
telescope to be $z_s=1.49$ (\cite{DICKINSON.1998}). Under an
assumption of spherically symmetric geometry, the projected lensing
mass within the arc radius, 9", is $M_{\rm lens}(<9") =
3.8\times10^{13}{\rm M_{\odot}}$. Here $\Omega_0=1$, $\Lambda=0$, and
$H_0=50\,{\rm km\,s^{-1}\,Mpc^{-1}}$ are adopted. The derived mass is
appropriate for clusters of galaxies rather than a single galaxy.

 A strong evidence for the existence of a galaxy cluster around
3C220.1 was obtained by the X-ray observation with ROSAT observatory
(\cite{HARDCASTLE_ETAL.1998}). The ROSAT HRI image revealed that the
the X-ray emission consists of a compact central component and an
extended component which can be attributed to cluster emission. The
ratio of the count rates of the compact to the extended components in
ROSAT HRI energy band is about 2 to 3.

In this paper we report the detection of an iron-K emission line with 
ASCA and discuss the origin of the X-ray emission.

\section{OBSERVATION AND RESULTS}

\subsection{Observation}
We observed 3C220.1 with the ASCA GIS and SIS for 40 ksec on 1998
April 11, during the AO6 period. The GIS was operated in the PH
nominal mode, while the SIS was in the Faint 1CCD mode. The data were
filtered by the standard ASCA data screening procedure.

X-ray emission centered at $(9^{\rm h} 32^{\rm m} 43^{\rm s}, +79^{\rm
d} 06^{\rm m} 39^{\rm s}) _ {\rm J2000.0}$ is detected.  The peak
position is about 0.2 arcmin off from the cataloged value of 3C220.1
but is consistent with it within the uncertainty of ASCA attitude
determinations (0.5 arcmin at 90\% confidence). The source count rates
are $(6.6 \pm 0.4)\times 10^{-3}$ counts/sec in the 0.7 -- 8 keV band
for the GIS and $(8.6 \pm 0.5)\times 10^{-3}$ counts/sec in the 0.5 --
8 keV for the SIS within a circle of 3' radius centered on the peak
position, after background subtraction. In the GIS and SIS fields,
there are several other bright sources in the vicinity of 3C220.1.
Most of them are not identified with known objects, except for one at
$(9^{\rm h} 31^{\rm m} 34^{\rm s}, +79^{\rm d} 04^{\rm m} 9^{\rm
s})_{\rm J2000.0}$ which is a radio-loud AGN (see comments in
\cite{HARDCASTLE_ETAL.1998}).

\subsection{Spectral Analysis}
Since the angular separations between 3C220.1 and the three of nearby
sources are 3.4 to 5.1 arcmin, the contamination of 3C220.1's energy
spectrum from these sources must be carefully treated. In order to
check the contribution due to the nearby sources, we have accumulated
the energy spectra in two different integration regions: (1) a
circular region centered at 3C220.1 with a radius of 3.0 arcmin and
(2) the same circular region as (1) but excluding three circular areas
centered on the nearby sources. The radii of the excluded regions are
proportional to the intensity of the nearby sources. We then evaluated
these two spectra by model-fittings. For both power-law and 
Raymond-Smith models, the resultant model parameters for the different
extraction regions are consistent with each other within the
statistical errors. In what follows, we show the results for case (1)
because at present only the azimuthally averaged response function is
available for the X-ray telescope; thus case (2) may involve some
systematic errors.

The PHA (Pulse Height Analysis) channels of the spectra were first
converted to PI (Pulse Invariant) channels and the spectra of the two
telescopes of the same system, i.e. SIS-0 and 1, and GIS-2 and 3,
respectively, were added together.  Some of the PI bins are combined
together so that the number of counts in any combined PI bin is
greater than 30.

We subtracted background spectra estimated in two different ways and
compared the results in order to estimate systematic errors.  In the
first case (a), the spectrum was estimated from the present
observation. For the GIS background, we accumulated events from
annular image regions whose centers are at the optical axes of the
telescopes. The outer and inner radii are respectively equal to the
maximum and minimum angular distances of the target spectrum region
from the optical axis. We excluded regions 4 arcmin from the target
and the contaminating sources. While for the SIS background, image
regions were selected outside 4 arcmin from the target and $\sim 3$
arcmin from the contaminating sources. In the other case (b), the
standard background spectrum which was obtained through blank
sky-field observations during the ASCA PV phase was used. The
background was derived from the same detector region as the target on
the SIS/GIS detector coordinates. The systematic errors of background
originate from spatial fluctuations of the Cosmic X-ray Background,
the detector-position dependence of Cosmic X-ray Background and
non-X-ray background, and temporal variation of non-X-ray
background. Their contribution to the above two backgrounds are
different. However, the spectral fits using the different backgrounds
provided consistent results within the statistical errors. Thus the
systematic errors are smaller than Poisson statistical errors. Since
the statistical errors of parameters using (b) are smaller than those
using (a) by 30 -- 50 \%, we will show the results obtained with (b)
hereafter.

\begin{figure}[h]
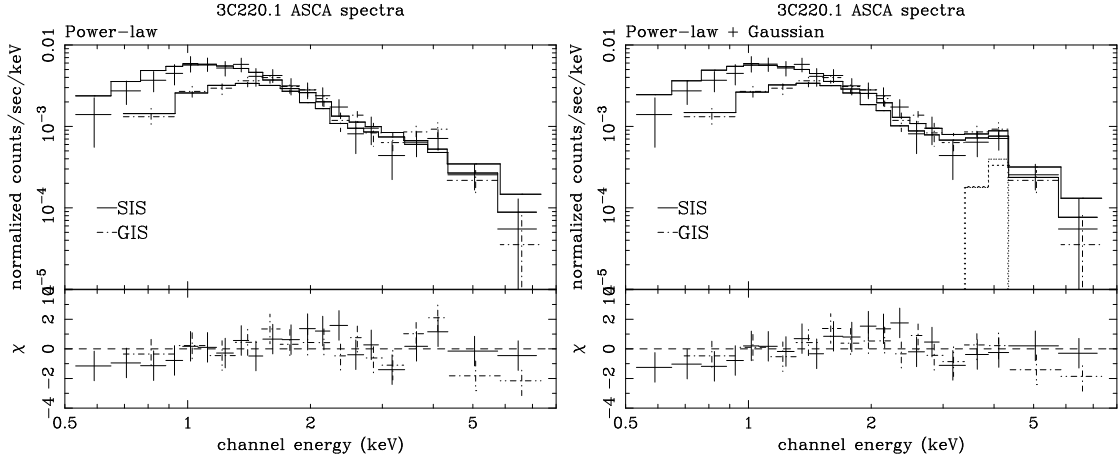

\centerline{
\psfig{figure=fig1.ps,height=6.0cm,angle=270}
\psfig{figure=fig2.ps,height=6.0cm,angle=270}
}
\caption{Spectral fits of ASCA SIS(0+1) and GIS(2+3) spectra with a
power-law model (left panel) and a power-law plus a Gaussian model
(right panel). The crosses denote the observed spectra and the step
functions show the best-fit model function convolved with the X-ray
telescope and the detector response functions. In the right panel, the
Gaussian component is shown with the dotted line. }

\label{fig:powerlaw}
\end{figure}

We fitted the GIS and SIS spectra simultaneously with model
spectra. First we tried a simple single component model; a single
power-law model with neutral absorption with a column density fixed at
the Galactic value, $N_{\rm H}=1.93\times10^{20}\,{\rm cm^{-2}}$
(\cite{STARK_ETAL.1992}). The absorption is fixed at this value
throughout this paper. The fits are acceptable with a best-fit
power-law photon index of 1.9 ($1.7-2.0$, 90\% error); however, the
fit leaves two excess data points at around 4 keV for both the SIS and
GIS spectra (Figure \ref{fig:powerlaw}).  \placefigure{fig:powerlaw}

The deviations of the data from the model in the 3.2 -- 4.3 keV band
are only 0.2 to 2.1$\sigma$ (Here $\sigma$ stands for a standard
deviation). However, the probability that we should observe such
deviations in any corresponding consecutive energy bins of two
different detectors is encouragingly low, $\sim 0.5$\%. We thus added
a narrow Gaussian emission line to the model with the Gaussian center
energy left free. The result is shown in Table \ref{tab:powerlaw} and
Figure \ref{fig:powerlaw}. In comparison to the single power-law model
fit, the fit improves from a $\chi^2$ value of 33.9 for 34 degrees
of freedom to 26.5 for 32 degrees of freedom. The improvement is
significant by the F-test at the 98.1\% confidence level. Thus we
conclude that the emission line feature is significant at the 98.1\%
confidence level.

The best-fit Gaussian center energy is 3.9 keV in our frame with the
90\% error range of $3.8-4.2$ keV. The most likely origin of this line
feature is a red-shifted iron emission line.  If we assume
low-ionization iron emission lines at 6.4 keV, the redshift is
estimated to be 0.63 ($0.53-0.69$, 90\% error), while if we assume 6.7
keV lines from helium-like iron, the redshift is 0.71 ($0.61-0.76$).
Therefore the redshift of the radio galaxy ($z=0.62$) is within the
error range in either case. Thus it is most likely that the emission
is originating from 3C220.1 or cluster surrounding it.
Then the central energy is estimated to be 6.3
($6.1-7.0$) keV in the rest frame. It can be interpreted either as a 6.4
keV low-ionization iron emission line which may be associated with
AGN, 6.7/6.9 keV lines from highly ionized irons which may be
attributed to cluster hot gas, or a combination of these two. We are
not able to distinguish these emission lines under the current limited
statistics and detector resolutions.
 
\begin{figure}[h]
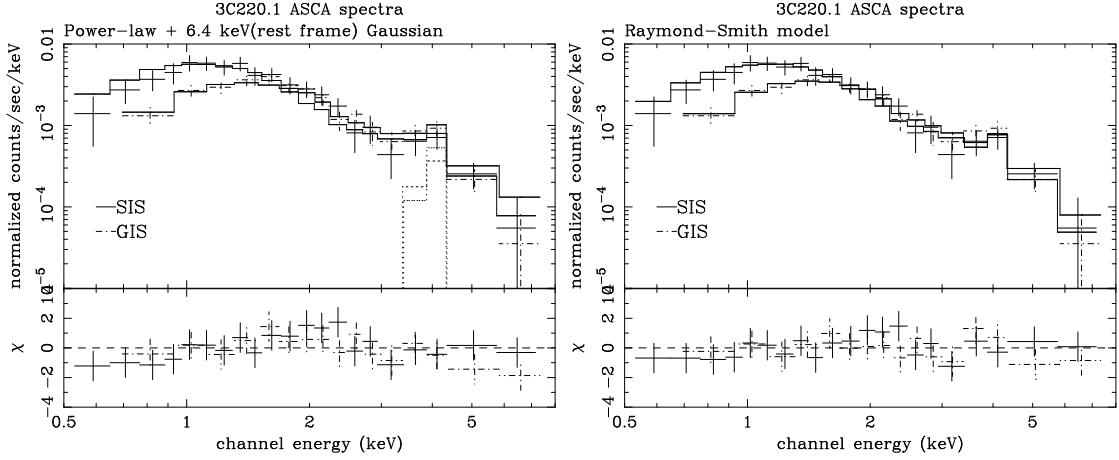

\centerline{
\psfig{figure=fig3.ps,height=6.0cm,angle=270}
\psfig{figure=fig4.ps,height=6.0cm,angle=270}
}
\caption{Spectral fits of ASCA SIS(0+1) and GIS(2+3) spectra with a
power-law plus 6.4 keV (rest-frame) Gaussian model (left panel), and
with a Raymond-Smith model (right panel). The redshift of the object
is assumed to be 0.62. In the left panel, the Gaussian component is
shown with the dotted line.  } \label{fig:spec} \end{figure}

Thus, we next performed fits with models corresponding to the above
two extreme cases: a power-law model plus a Gaussian emission line
with the line center energy fixed at 6.4 keV, and a Raymond-Smith
model representing an optically thin thermal plasma emission. We fixed
the redshift value at the position of the radio galaxy, 0.62. The
results of these fits are shown in Figure \ref{fig:spec} and Table
\ref{tab:result}, where the model parameters for the GIS and SIS
spectrum are combined.  
Although both models are acceptable at the 90\% confidence limit, the
Raymond-Smith model gives a smaller reduced $\chi^2$ value. The
luminosity in the $2-10$ keV band is $1\times10^{45}$ erg/s, assuming
a distance of $z=0.62$.

\begin{figure}[h]
\centerline{
\psfig{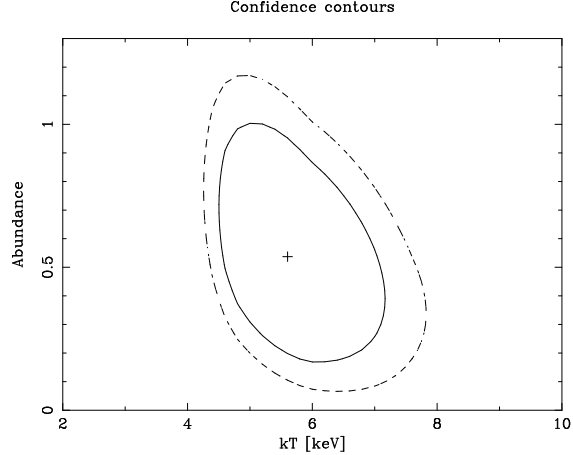}
}
\caption{$\chi^2$ contours of the Raymond-Smith fit. The 90\%
single-parameter and two-parameter error domains are shown as contours
with solid and broken curves, respectively. 
The position of the $\chi^2$ minimum is 
denoted with a cross.}
\label{fig:contour} 
\end{figure}

For the non-thermal model consisting of a power-law plus a 6.4keV
line, we obtained an equivalent width of $\sim500$ ($190-780$) eV,
while for the Raymond-Smith fit, we found a metal abundance of 0.54
($0.17-1.0$) solar. In Figure \ref{fig:contour} we have plotted the
$\chi^2$ contours as a function of two of the Raymond-Smith model
parameters: k$T$ and metal abundance.

\placefigure{fig:spec}
\placefigure{fig:contour}

\section{DISCUSSION}
We have detected an emission line at 3.9 keV (3.8 -- 4.2 keV) in our
rest frame, which corresponds to 6.1 -- 7.0 keV at $z=0.62$.  Within
the present statistics, the observed spectra are consistent with two
different models; a non-thermal model which consists of a power-law
continuum and a 6.4 keV emission line, and a Raymond-Smith thermal
model with temperature about 6 keV. In this section, we discuss the
origin of the X-ray emission.

The radio source 3C220.1 is classified as an FRII narrow emission line
galaxy (NELG). \cite{TURNER_ETAL.1997} investigated the narrow iron K
line at 6.4 keV from type 2 AGNs systematically, and reported that
most of the NELG show an equivalent width smaller than 200 eV and on
average $\sim$ 100 eV. Thus for 3C220.1, the derived equivalent width
under a non-thermal model is a factor of 2 to 8 larger than the
typical NELGs if one attributes all of the line intensity to the radio
galaxy 3C220.1.  This indicates a large ($ > 50 $\%) fraction of the
iron line is emitted from the other emission region: most likely the 
intracluster medium in the galaxy cluster.

The ROSAT HRI observations revealed that the X-ray emission consists
of a compact component and an extended component which is
significantly extended more than 10 arcsec in radius and carries
about 60\% of the HRI photons (\cite{HARDCASTLE_ETAL.1998}).
Since the two spectral models in our analysis are not distinguished
within the statistics, any combination of the two models should also
be statistically acceptable.  If the compact component carries 40\% of
the photons in the ASCA energy band and contains an iron emission line
of 100 eV equivalent width in its own spectrum, the equivalent width
of the rest of the emission is expected to be 250 eV to 1200 eV. This
value is appropriate for thin thermal emission of k$T\sim6$ keV with a
metal abundance of 0.2 to 1.6 solar.

If we consider about 60\% of the total emission to arise from the
cluster, the luminosity and the temperature obtained from the spectral
fit are consistent with the luminosity-temperature relation for nearby
clusters obtained by \cite{DAVID_ETAL.1993} within the scatter of data
points.

The masses of distant clusters determined from the gravitational arc
(lens mass) and determined from the X-ray observations (X-ray mass)
have been compared by several authors
(e.g. \cite{WU_FANG.1997}). These results show that in general the
X-ray mass is either consistent with or smaller than the lens mass.
Under the assumption of hydrostatic equilibrium, isothermal and
spherical distribution of the intracluster gas, and $\beta$-model
surface brightness distribution, the X-ray mass projected within a
radius $r$ is estimated as 
$ M_{\rm x \beta}(r) = ( 3 {\rm k} T \beta/{\rm G} \bar{m})
                     (\pi/2)( r^2/ \sqrt{r^2 + r_c^2}) $ ,
 where k is Boltzmann constant and $\bar{m}$ is the mean mass per
plasma particle (\cite{OTA_ETAL.1998}). Adopting the best-fit
$\beta$-model parameters obtained from the ROSAT HRI observations
($\beta=0.9$ and the core radius, $r_c=13"$), we find that the X-ray
mass contained within the arc radius is equal to or smaller than the
lens mass if the temperature of the X-ray emission is lower than 6.4
keV. The determined X-ray temperature of $5.6^{+1.5}_{-1.1}$ keV from
the single Raymond-Smith model fits is consistent with this.

In conclusion, a large fraction of the iron line emission and the
continuum emission associated with the line is likely to originate
from the intracluster medium in the galaxy cluster around
3C220.1. This is strong evidence for the existence of a cluster of
galaxies including 3C220.1.

\acknowledgments 

We are very grateful to M. Dickinson for valuable information and
discussions on the recent HST results and also to D.M. Worrall for
sending us the ROSAT results before publication. We would like to
thank D. Audley for his careful review of the manuscript. This work
was supported in part by the Japan Society for the Promotion of
Science.

\newpage

{\bf Figure Captions}\\

\newpage

\begin{table}
\begin{center}
\caption{Results of Spectral Fits}
\begin{tabular}{lll}\hline\hline
 Model & Parameter & Value ( error\tablenotemark{\dagger} ) \\\hline 
 Power-law  &Photon Index & 1.9 ($1.8-2.1$)\\ 
	    &$\chi^2$/d.o.f&33.9/34\\\hline 
 Power-law plus Gaussian &Photon Index & 1.9 ($1.8-2.1$)\\ 
		& Gaussian Center Energy [keV] & 3.9 ($3.8-4.2$)\\
		&$\chi^2$/d.o.f&26.5/32\\\hline 
 \end{tabular}
\label{tab:powerlaw}
\end{center}
\tablenotetext{}{The absorption column density was fixed at the
Galactic value; $N_{\rm H}=1.93\times10^{20}[{\rm cm^{-2}}]$.
The sigma value of the Gaussian functions is fixed at 0.03 keV.} 
\tablenotetext{\dagger}{The quoted errors correspond to a single parameter error at 90\% confidence.}  
\end{table}

\begin{table}
\begin{center}
\caption{Results of Spectral Fits}
\begin{tabular}{lll}\hline\hline
 Model & Parameter & Value ( error\tablenotemark{\ast} ) \\\hline 
 Power-law plus Gaussian&Normalizaion \tablenotemark{\dagger}
						& 1.3 ($1.1-1.4$)\\
			&Photon Index & 1.9 ($1.8-2.1$)\\ 
			&Equivalent width [eV]\tablenotemark{\ddagger}
						& 480 ($190-780$)\\
			&$\chi^2$/d.o.f&26.7/33\\
			&$L_{\rm X}^{2-10}$ [erg/s]\tablenotemark{\clubsuit}
						&$1.1\times10^{45}$\\\hline 
 Raymond-Smith	&Normalizaion \tablenotemark{\dagger}& 9.4 ($8.1- 10.7$)\\
		&k$T$[keV] & 5.6 ($4.5-7.1$)\\ 
		&Abundance [Z$_{\odot}$] &0.54 ($0.17-1.0$)\\
		&$\chi^2$/d.o.f&18.1/33\\
		&$L_{\rm X}^{2-10}$ [erg/s]\tablenotemark{\clubsuit}
						&$1.0\times10^{45}$\\\hline 
\end{tabular}
\label{tab:result}
\end{center}
\tablenotetext{}{The absorption column density was fixed at the
Galactic value; $N_{\rm H}=1.93\times10^{20}[{\rm cm^{-2}}]$.} 
\tablenotetext{\ast}{The quoted errors correspond to a single parameter error at 90\% confidence.}  
\tablenotetext{\dagger}{Flux at 1keV (in $10^{-4}$ photons s$^{-1}
$keV$^{-1}$ cm$^{-2}$) for the power-law fit or $\int n_{e} n_{h}
dV/4\pi D^2$ (in $10^{-18}$ cm$^{-5}$) for the Raymond-Smith fit,
where $D$ is the distance to the source (cm), $n_e$ and $n_h$ are the
electron and hydrogen densities (cm$^{-3}$) .}
\tablenotetext{\ddagger}{A narrow line at 6.4 keV is assumed.}
\tablenotetext{\clubsuit}{Absorption-corrected $2-10$ keV luminosity
assuming a distance corresponding to $z=0.62$.}

\end{table}


\begin{thebibliography}{}
\bibitem[{{David}{ \it et~al.}}{ (1993)}]{DAVID_ETAL.1993}
{David}, L. P., {Slyz}, A., {Jones}, C., {Forman}, W., \& {Vrtilek}, S. D., 
 1993, \apj, {412}, 479.

\bibitem[{{Dickinson}}{ 1984}]{DICKINSON.1984}
{Dickinson}, M., 1984, 
\newblock Ph.D. thesis, University of California.

\bibitem[{{Dickinson}}{ 1998}]{DICKINSON.1998}
{Dickinson}, M., 1998, 
\newblock private communication.

\bibitem[{{Fort} \& Mellier}{ 1994}]{FORT_MELLIER.1994}
{Fort}, B., \& {Mellier}, Y., 1994, \aapr, {5}, 239.

\bibitem[{{Hardcastle}{ \it et~al.}}{ 1998}]{HARDCASTLE_ETAL.1998}
{Hardcastle}, M. J., {Lawrence}, C. R., \& {Worrall}, D. M., 1998,
\apj, {504}, 743.

\bibitem[{{Ota}{ \it et~al.}}{ 1998}]{OTA_ETAL.1998}
{Ota}, N., {Mitsuda}, K., \& {Fukazawa}, Y., 1998, \apj, {495}, 170.

\bibitem[{{Stark}{ \it et~al.}}{ 1992}]{STARK_ETAL.1992}
{Stark}, A. A., {Gammie}, C. F., {Wilson}, R. W., {Bally}, J., {Linke}, R.A., 
{Heiles}, C., \& {Hurwitz}, M., 1992, \apjs, {79}, 77.

\bibitem[{{Turner}{ \it et~al.}}{ (1997)}]{TURNER_ETAL.1997}
{Turner}, T. J., {George}, I. M., {Nandra}, K. \& {Mushotzky}, R. F., 1997, 
\apjs, {113}, 23.

\bibitem[{{Wu} \&  Fang}{ 1997}]{WU_FANG.1997}
{Wu}, Z-P. \& {Fang}, L-Z, 1997, \apj, {483}, 62.
\end{thebibliography}
\end{document}